\documentclass[3p,times]{elsarticle}

\usepackage{amssymb}

\usepackage[figuresright]{rotating}

\begin{document}

\begin{frontmatter}




\title{Chaotic interaction between thermodynamic systems}

 \author[label1,label2,label3]{Ekrem Aydiner}
 \address[label1]{Theoretical and Computational Physics Research Laboratory, Istanbul University, 34134 İstanbul, Türkiye}
\address[label2]{Department of Physics, İstanbul University, Fatih, 34134 İstanbul, Türkiye}
\address[label3]{Department of Physics, Koç University, Sarıyer, 34450, İstanbul, Türkiye}

\address{}

\begin{abstract}
In previous study \cite{Aydiner_2024}, we proposed a new physical law applicable to both particle and thermodynamical systems. Additionally, we introduced a physical definition of chaos and self-organization. In the present work, we extend this novel interaction scheme to various thermodynamical systems in order to test the universality of the proposed law. Our results demonstrate that the dynamics of these systems also exhibit chaotic behavior under different conditions. We discussed the relationship between chaotic dynamics and the least action, and for the first time, we proposed the \textit{Chaotic Action Principle} (CAP). Furthermore, we examined the validity of the physical and philosophical foundations of this new law and the associated definition of chaos. Finally, based on CAP, we proposed a new conceptual definition of complexity.
\end{abstract}


\end{frontmatter}



\section{Introduction} \label{intro}

The scientific revolution began in the 17th century with  Newton's laws of classical mechanics and the development of calculus \cite{Newton_1687tb}. Later, inspired by Newton, Maupertuis proposed the \textit{Least Action Principle} (LAP) instead of the idea of the \textit{least path} and \textit{least time} \cite{Maupertuis_1744}. This principle states that a system evolves in such a way as to extremize a quantity known as the action, which is related to the system's energy. Today, we know that the LAP has played a central role in the development of physics.
For example, historical figures such as Euler \cite{Euler_1767tb}, Lagrange \cite{Lagrange_1772tb}, Hamilton \cite{hamilton1834xv,hamilton1835second}, and Jacobi \cite{jacobi1837theorie}, drawing upon the LAP established classical mechanics by reformulating Newtonian mechanics within a different formalism. In the same \textit{classical period}, another prominent historical figure who expanded upon Newton's work, Laplace, argued that the universe is governed by a clear and deterministic order \cite{Laplace1829}.  Indeed, Newtonian physics is both linear and deterministic. In other words, within the framework of Newtonian physics, the future behavior of a system can be predicted with complete certainty, provided that its initial conditions are known. However, later, it was understood that numerous physical, biological, chemical, social, economic and climatic systems in nature exhibit non-linear and non-deterministic behavior, contrary to the Newtonian and Laplacian worldview.

Indeed, when Poincaré \cite{Poincare_1890tb} examined the three‑body problem in 1890, he discovered that its solutions could be both unstable and unpredictable—an insight that, about a century later, would give rise to chaos theory. In the ensuing decades, the field was advanced by seminal contributions from Hadamard \cite{Hadamard_1898,Gutzwiller_1980}, Sharkovskiĭ \cite{Sharkovskii1964}, van der Pol \cite{Balth1926}, Kolmogorov \cite{Kolmogorov1954,Arnold1997,Moser1962}, Lorenz \cite{Lorenz_1963}, May \cite{May1976simple}, Yorke \cite{Yorke1975}, Ruelle \cite{Ruelle1971}, Julia \cite{Julia_1918}, Mandelbrot \cite{Mandelbrot1973formes}, and Feigenbaum \cite{Feigenbaum1978quantitative}. These developments profoundly challenged the Newtonian paradigm: rather than the strictly linear and deterministic universe envisaged by Laplace and Newton, many physical system in the nature can show itself as inherently non-linear and chaotic.

However, now, despite all these important developments in the field of chaos, it can be said that the relationship between the known \textit{order of nature} and \textit{chaos} is not yet fully understood. At this point, various questions can be asked. For example: \textit{What do non-linear dynamics in nature and chaos in general represent?} \textit{Is chaos a simple example that only occurs in some special cases?} \textit{Is nature composed of linear and non-linear systems or is it completely non-linear?} \textit{If nature is completely non-linear, doesn't this contradict our knowledge so far?} \textit{How is order established in non-linear nature?} Even if it is not possible to answer all of these questions, the existence of an uncountable number of chaotic systems in nature suggests that there must be a relationship between chaos and the order of nature. On the other hand, another known fact is that nature is covered with fractal geometry \cite{Mandelbrot_1982,Feder_1988,Barnsley_1993,Peitgen_2004}. Indeed, this situation may imply that behind the fractal structure of nature and the order of the universe there is chaotic action \cite{Aydiner_2024}. Therefore, it is impossible to ignore these interesting metaphors to understand the hidden order behind the visible world. This idea is not the first. There are also researchers who imply that nature as a whole behaves chaotically \cite{Linde_1987,Pietronero1987,Guth_2007,Ambjorn_2005,Connes_1994,Palmer2021}. As it is known, the three great theories of 20th century physics; general relativity, quantum and chaos theory are incompatible with each other. For instance, in 2021, Palmer proposed that quantum and general relativity could be unified through chaos theory \cite{Palmer2021}. Because of this importance of the subject, all these physics-originated philosophical questions point to the need for a new perspective that would not exclude known physics but would also offer a more general perspective. This is the main motivation for this work.

With the motivation to develop a new perspective, we put forward the idea that nature may behave completely chaotic in our previous work \cite{Aydiner_2024}. Based on the new interaction scheme we proposed, we showed that the dynamics of two interacting thermodynamic systems must be chaotic. We generalized our findings and proposed a new law based on the idea that the interactions of dynamic and thermodynamic systems must have a universal property, and we gave a physical definition for the first time in the literature by establishing a connection between chaotic dynamics and the LAP. The dynamic systems law and the definition of chaos we proposed are as follows, respectively:

\textit{--"The dynamics of all coupled interacting particle and thermodynamic systems are chaotic."} 

\textit{--"Chaos is the minimum action of the coupled, synchronized or self-organized interacting systems."} 

In the previous study \cite{Aydiner_2024}, we considered two interacting thermodynamic systems. We assumed that these systems interact via heat and particle exchange. We modeled this interaction in accordance with the new interaction scheme we proposed, the mutual and self-interaction schemes, and obtained the interaction equations. We showed that the interaction dynamics of these two thermodynamic systems under conditions $\delta Q \ne 0$, $\delta N \ne 0$ and $dV=0$ are chaotic. 

In this study, our first aim is to show that the physical law we propose is universal. For this purpose, in this study, we will try to extend the discussion to different thermodynamic systems which interact under different conditions: i) $\delta Q \ne 0$, $\delta N=0$ and $dV=0$ and ii) $\delta Q \ne 0$, $\delta N\ne 0$ and $dV \ne0$. Using the same interaction schema in~\cite{Aydiner_2024}, we will show that the interactions between these systems are chaotic. Second, we will discuss why the physical definition of chaos is correct by taking into account previous research and discuss the correctness of the definition of chaos by trying to reveal the philosophical connection between the definition of chaos and the minimum action principle. Third, to introduce the principle of chaotic action by generalizing the principle of least action based on the idea that chaos minimizes action. Fourth, to discuss how micro and macro systems in nature behave in accordance with the principle of chaotic action. Five, to initiate a discussion on the concept of complexity. The last four items we have listed here will be discussed in the discussion section of the study.

This work is organized as follows: In Section \ref{model}, we briefly introduce the interaction model based on Ref. \cite{Aydiner_2024}. In Section \ref{caseI}, we consider a case for $\delta Q \ne 0$, $\delta N=0$, and $dV=0$. In Section \ref{caseII}, we consider a case for $\delta Q \ne 0$, $\delta N\ne 0$, and $dV \ne0$. In both sections, we introduce new interaction equations for two cases, and we will analyze the phase space and chaotic behaviours of the interacting systems. In Section \ref{disc}, we will discuss the new results obtained and the physical and philosophical aspects of the proposed laws and definitions. In addition, in this section, we will also introduce a new action principle and definition of the complexity. Finally, in the last Section \ref{conc}, conclusions are presented.

\section{Revisit New interaction schema} \label{model}

\begin{figure} [h!]
	\centering
	\includegraphics[scale=0.75]{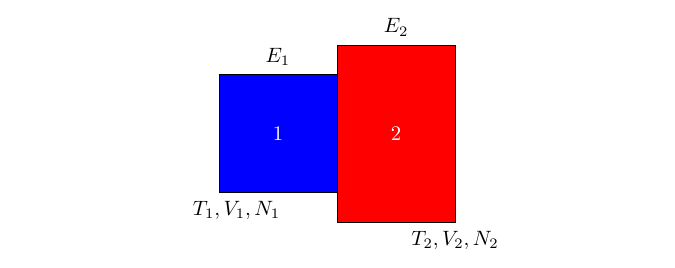}
	\caption{Two thermodynamic systems in contact. $E_{1,2}$ are the energies of the systems, $T_{1,2}$ denote the temperatures, $V_{1,2}$ represents the volumes, and $N_{1,2}$ are the number of particles.}
	\label{cXfig1}
\end{figure}

Let us imagine that two separate thermodynamic systems with different energies, temperatures, volumes, and numbers of particles are in contact with each other, as in Fig.\ref{cXfig1}. According to the traditional thermodynamic approach, if these two systems are not in equilibrium, that is, for example, $T_{1} \ne T_{2}$, $p_{1} \ne p_{2}$, or $\mu_{1} \ne \mu_{2}$, then the systems are far from equilibrium and will continue to interact with each other until they reach equilibrium. Changes in internal energy in each systems can be derived from Eq.(\ref{dU}): 
\begin{eqnarray} \label{dU}
	dU = \delta Q  + \sum_{i=1}^{N} \delta W_{i}
\end{eqnarray}
where $dU$ corresponds to energy exchange in the system, $\delta Q$ is heat changes, and $\delta W_{i}$ denotes work which is done on the system. The energy changes in each system can be obtained by using the expression given in Eq.(\ref{dU}), also known as the law of conservation of energy.

\begin{figure} [h!]
	\centering
	\includegraphics[scale=0.75]{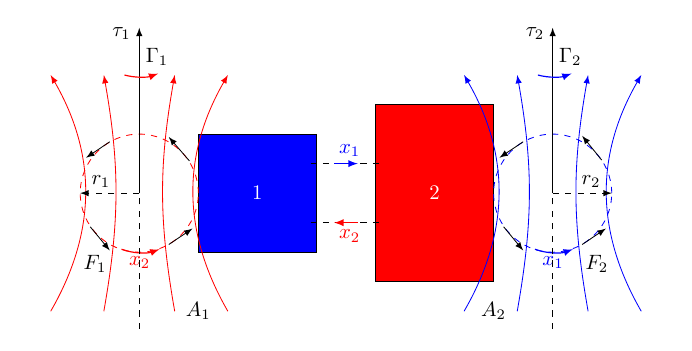}
	\caption{Schematic representation of two thermodynamic systems interacting in contact. The vectorial attractor fields and torque behavior on the self-interacting loops. $\Gamma_{1,2}$ denotes non-holonomic variables in Eq.(\ref{self-int}). $\vec{A}_{1,2}$ are the attractor vectorial fields caused by driven force $\vec{F}_{1,2}$ and $\vec{\tau}_{1,2}$ correspond to torque on the loops. These variables are defined in Ref.~\cite{Aydiner_2024}.}
	\label{cXfig2}
\end{figure}

However, according to the new interaction schema, when two thermodynamic systems interact, two inseparable and interdependent interactions occur. These are mutual interaction and self-interaction. As stated in the previous article, mutual interaction represents the energy exchange between two systems, while self-interaction is the type of interaction that results from the mutual interaction and acts on each system itself. Mutual interaction and self-interaction are given in Fig.\ref{cXfig2}. In this figure, the mutual energy interaction between system 1 (blue box) and system 2 (red box) is shown as $x_{1}$ and $x_{2}$ between the two systems. Self-interaction loops are also shown on the left of system 1 and on the right of system 2. Although the loops are drawn outside the systems, they actually act on the system. In this model, mutual and self-interactions continue due to the existence of the classical or quantum fluctuation around the systems equilibrium.

It should be noted that this new self-interaction scheme includes hidden variables and hidden vectorial fields as seen in Fig.\ref{cXfig2}. According to the model, these hidden variables and hidden vectorial fields lead to dynamics in self-interaction loops. It can be seen that different hidden attractors have been proposed in the literature. For example, the concept of hidden attractors was first introduced in connection with Hilbert’s 16th problem formulated in 1900 \cite{Hilbert1902}. Similarly, Kuznetsov et al. reported the chaotic hidden attractor in Chua’s circuit \cite{Leonov2013,leonov2012hidden,leonov2015homoclinic}. This research has since led to further efforts to locate hidden attractors in various examples \cite{Leonov2011,Dudkowski2015,Kuznetsov2014,kumarasamy2023,Kuznetsov2023,jafari2015recent}.

Now let's look at how the mutual interaction and self-interaction mechanisms are established. According to this scheme the energy exchanges for two coupled interacting systems can be given in the form~\cite{Aydiner_2024}:
	\begin{eqnarray}  \label{dU_S1}
		dU_{1} (x_{2} )= \delta Q_{1}(x_{2})  + \delta W_{1}(x_{2})  + \delta N_{1}(x_{2})
	\end{eqnarray}
	\begin{eqnarray} \label{dU_S2}
		dU_{2}(x_{1}) = \delta Q_{2}(x_{1}) + \delta W_{2}(x_{1})  + \delta N_{2}(x_{1})
	\end{eqnarray}
where $	dU_{1} (x_{2} )$ and $	dU_{2} (x_{1} )$ denote the energy exchanges on the system 1 (blue box) and 2 (red box), respectively. Here, $x_{1}$ is considered the energy flux passing through system 1, and $x_{2}$ is considered the energy flux passing through system 2. 

The idea behind this interaction model is to write the energy change in a system in a source-dependent form. This approach allows us to write the interaction energy conservation equations in non-equilibrium form without violating energy conservation. In this scheme internal energy exchanges between systems will be conserved:
\begin{eqnarray} \label{energy-cons-U}
	\oint dU(x_{i}) = 0
\end{eqnarray}
where $i=1,2$.
At the same time, total energy is constant:
\begin{eqnarray} \label{energy-cons-E}
	\sum_{i=1}^{2} E_{i} = const
\end{eqnarray}

The mutual interactions can be obtained from Eq.(\ref{dU_S1}) and Eq.(\ref{dU_S2}). The details are presented in the previous work \cite{Aydiner_2024}. To avoid repetition, we will not give the details of the derivation here. However, we will give the mutual interaction expressions for the two models directly in Eq.(\ref{case1-eq}) and Eq.(\ref{Aydiner2}). Similarly, the self-interaction equations can be also obtained by using variables and mechanism given on the loops in Fig.\ref{cXfig2}. These equation remain as given in the previous work. They are:
\begin{equation} \label{self-int}
	\frac{d \Gamma_{1}}{dt} = 1 - x_{1} x_{2} ,    \quad
	\frac{d \Gamma_{2}}{dt} =  1 - x_{1} x_{2}    
\end{equation} 
or for solid systems that are assumed not to rotate while in contact, so Eq.(\ref{self-int}) can be written as
\begin{eqnarray} \label{self-int-solid}
	\frac{d \Gamma_{1}}{dt} = - x_{1} x_{2} ,    \quad
	\frac{d \Gamma_{2}}{dt} =  - x_{1} x_{2}    .
\end{eqnarray}

To sum up, the interaction scheme summarized here can be applied to two different systems, namely $\delta Q \ne 0$, $\delta N=0$, $dV=0$ and $\delta Q \ne 0$, $\delta N\ne 0$, $dV \ne0$. Below, we will examine these two cases in detail and discuss the results. On the other hand, as we stated in the introduction, we will expand the discussion in Section \ref{disc} by taking into account the results obtained in this study.

\section{Case I: $\delta Q \ne 0$, $\delta N=0$ and $dV=0$} \label{caseI}

For the case $\delta Q \ne 0$, $\delta N=0$ and $dV=0$ the mutual interaction equation as 
\begin{equation}  \label{case1-eq}
	\frac{d x_{1}}{dt}  =  \Gamma_{1} x_{2}, \qquad
	\frac{d x_{2}}{dt}  =  \Gamma_{2} x_{1}  
\end{equation}
where $\Gamma_{1,2}$ are the non-holonomic variables which are denote \textit{energy transfer velocity}.

For Case I, by using Eqs.(\ref{self-int}) and (\ref{case1-eq}), we obtain interaction equation for the coupled and interacting two canonical ensembles i.e. thermodynamic systems as
\begin{eqnarray}  \label{Aydiner1}
	\frac{d x_{1}}{dt}  &=& x_{2} x_{3}   \nonumber \\
	\frac{d x_{2}}{dt}  &=&  (x_{3}- q)  x_{1}  \\
	\frac{dx_{3}}{dt} &=& 1 - x_{1} x_{2} \nonumber
\end{eqnarray} 
where  $q$ is the control parameters $q=\Gamma_{1}-\Gamma_{2}$. Eq.(\ref{Aydiner1}) has a strongly non-linear form. Here, $x_{1}$ and $x_{2}$ indicate energy exchange resulting from the heat exchange. In this example, there is only heat exchange between the two systems. 
In other words, there is no particle exchange between systems 1 and 2. Also, since the volume does not change in both systems, no work is done on the system due to the volume change. Similarly, the self-interaction affecting systems 1 and 2 is also safely dependent on heat.
\begin{figure} [h!]
	\centering
	\includegraphics[height=6cm,width=7cm]{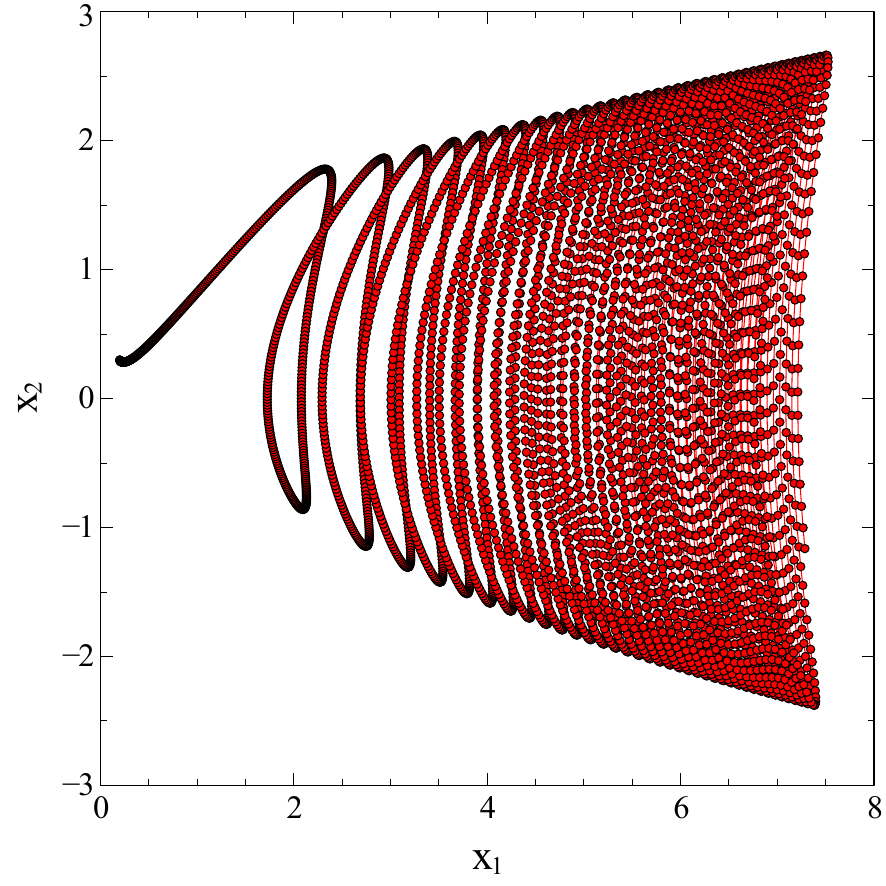}
	\caption{Non-periodic orbital in $x_{1}-x_{2}$ plane for $q=3.46$ and attractor basin.}
	\label{phase-1}
\end{figure}
\begin{figure} [h!] \centering \includegraphics[height=6cm,width=7cm]{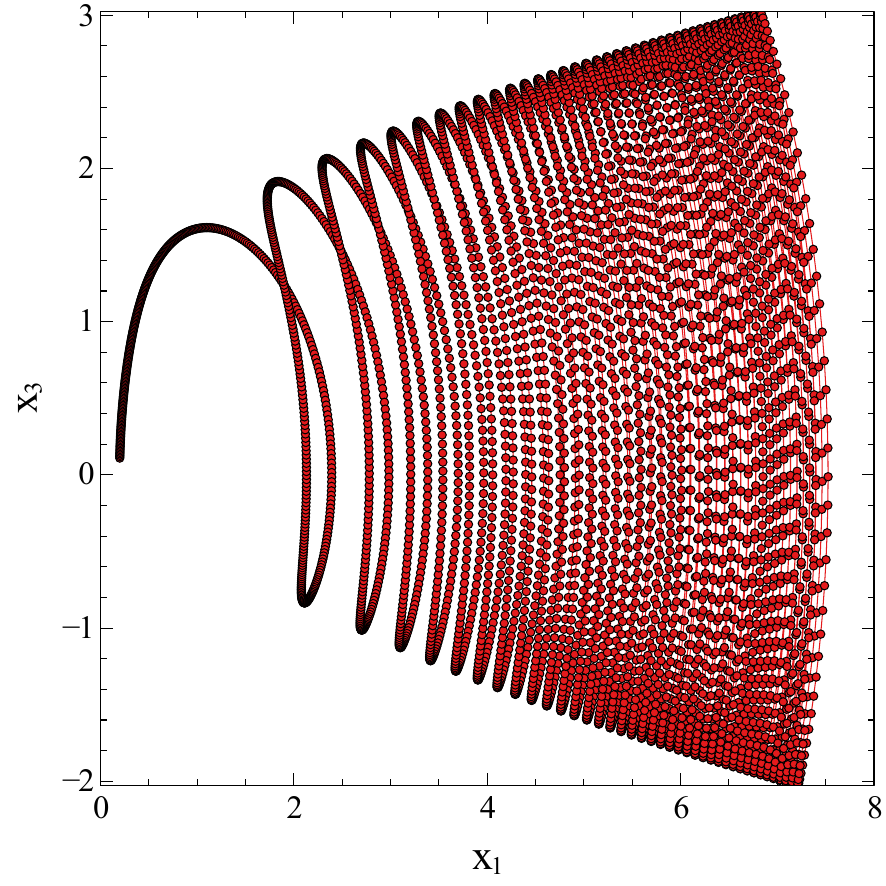}	\caption{Chaotic attractor and trajectory in $x_{1}-x_{3}$ plane for $q=3.46$. }\label{phase-2}	
\end{figure}
\begin{figure} [h!]	\centering	\includegraphics[height=6cm,width=7cm]{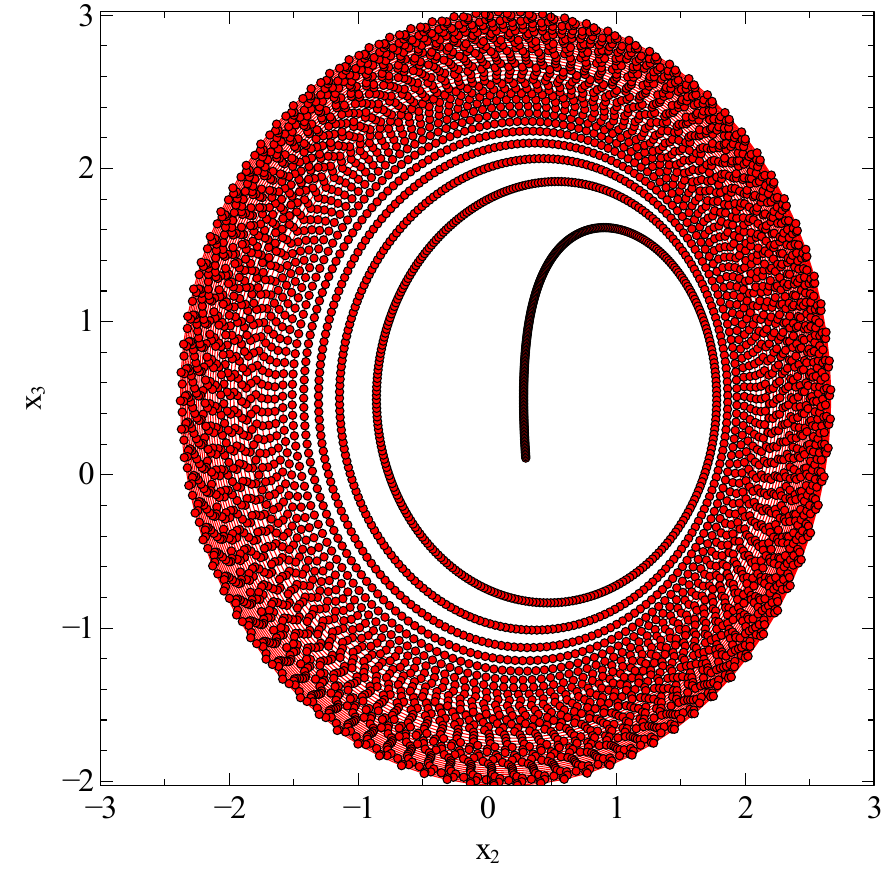}\caption{Chaotic attractor and trajectory in $x_{2}-x_{3}$ plane for $q=3.46$.}\label{phase-3}
\end{figure}

\begin{figure} [h]
	\centering
	\includegraphics[height=6cm,width=7cm]{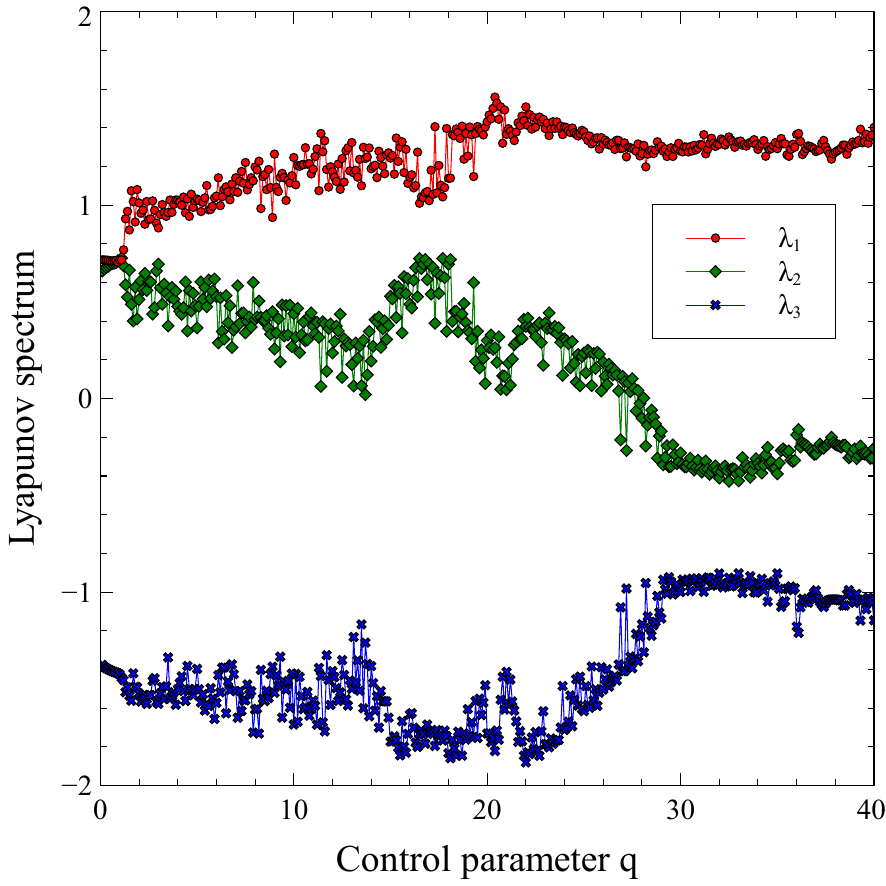}
	\caption{Three Lyapunov exponents versus control parameter $q$. }
	\label{Lyap-1}
\end{figure}

Using the Runga-Kutta method and the linearized algorithm \cite{Wolf1985}, we numerically solved Eq.(\ref{Aydiner1}) with FORTRAN 90. We obtained phase-space solutions and Lyapunov exponents belonging to the coupling equations. Phase trajectories and Lyapunov exponents are given in Figs.\ref{phase-1}-\ref{phase-3} and Fig.\ref{Lyap-1}, respectively. 

The solutions of $x_{1}$ and $x_{2}$ for the control parameters $q=3.46$ are given in Fig.\ref{phase-1}. 
It is clearly seen that the trajectory in the $x_{1}-x_{2}$ plane does not intersect and repeat in the attractor basin. One can see that the attractor appears around $x_{2}=0$ and the trajectory proceeds in the $x_{1}$ direction.

Similarly, solutions of $x_{1}$ and $x_{3}$  for the control parameters $q=3.46$ are given in Fig.\ref{phase-2}, 
the attractor appears in the $x_{1}-x_{3}$ plane around $x_{3}=0$. One can see that orbit does not make periodic changes and moves in the $x_{1}$ direction without cutting itself. In fact, the attractor behaviours in Fig.\ref{phase-1} and Fig.\ref{phase-2} can be clearly seen in Fig.\ref{phase-3}. In fact, the attractor and the trajectory appear around $x_{2}=0$ and $x_{3}$ on $x_{2}-x_{3}$ in Fig.\ref{phase-3}. 

Additionally, we have tested that the character of the phase-space solution does not change for the various running step time. One can see that the chaotic attractor is located at only around one fixed point. On the other hand, the phase-space solution in Figs.\ref{phase-1}, \ref{phase-2} and \ref{phase-3} are plotted for only same arbitrary $q$ value. However, we see that the character of the phase-space solutions do not change for the different $q$ values. 
One can see that from Figs.\ref{phase-1}, \ref{phase-2} and \ref{phase-3} the trajectories within the
attractors do not intersect. If the trajectories were to
intersect, the solutions would no longer be expected to
exhibit chaotic behavior. The crucial point here is the
demonstration of the existence of chaotic attractors. It
is possible to reproduce such plots for various $q$ values.
Whether these attractors are indeed chaotic can be further clarified using the Lyapunov exponent.

Now we discuss the Lyapunov \cite{Lyapunov1907} spectrum of the system in Eq.(\ref{Aydiner1}). Positive Lyapunov exponents are known to indicate chaotic behaviour of the system. We obtained the Lyapunov spectrum using the Wolf algorithm and plotted it in Fig.\ref{Lyap-1}. 
One can see that the system produces three positive Lyapunov exponents in this interaction schema. As seen in Fig.\ref{Lyap-1}, one of the exponents which receive a red curve is always positive for all values $q$. However, the other two Lyapunov exponents (yellow and blue curves) take negative and positive values for the different $q$ values. It is strongly provided that the dynamics of the two coupled and interacting canonical or close thermodynamic systems are chaotic like two grand canonical systems \cite{Aydiner_2024}.

\section{Case II: $\delta Q \ne 0$, $\delta N\ne 0$ and $dV \ne0$} 
\label{caseII}

In this section, we will also consider volume changes. As a result of the interaction, the volume of the systems may change, and this prevents work on the system. The work performed on the system is defined as $\delta W = -p dV$. In this case, for Case II, i.e., for  $\delta Q \ne 0$, $\delta N \ne 0$ and $dV \ne 0$, the equations Eq.(\ref{Aydiner1}) can be written in the following form:
\begin{eqnarray}  \label{Aydiner2}
	\frac{d x_{1}}{dt}  &=& x_{2} x_{3} -  x_{1}  - \int_{0}^{\tau} x_{1}(t) dt  \nonumber \\
	\frac{d x_{2}}{dt}  &=&  (x_{3}- q)  x_{1}  - x_{2} -  \int_{0}^{\tau} x_{2}(t) dt  \\
	\frac{dx_{3}}{dt} &=& 1 - x_{1} x_{2} \nonumber
\end{eqnarray} 
Eq.(\ref{Aydiner2})
has also strongly non-linear and which describes the dynamics of the
interacting systems via heat  and particle transfer. Here, $x_{1}$ and $x_{2}$ refer to the energy exchange resulting from heat and particle exchange. Similarly, the self-interaction  act on the systems 1 and 2 also depends on heat and particle exchange. In this example, since the volume changes in both systems, work is done on the system due to the volume change.

\begin{figure*} [ht]
	\centering
	\includegraphics[height=6cm,width=7cm]{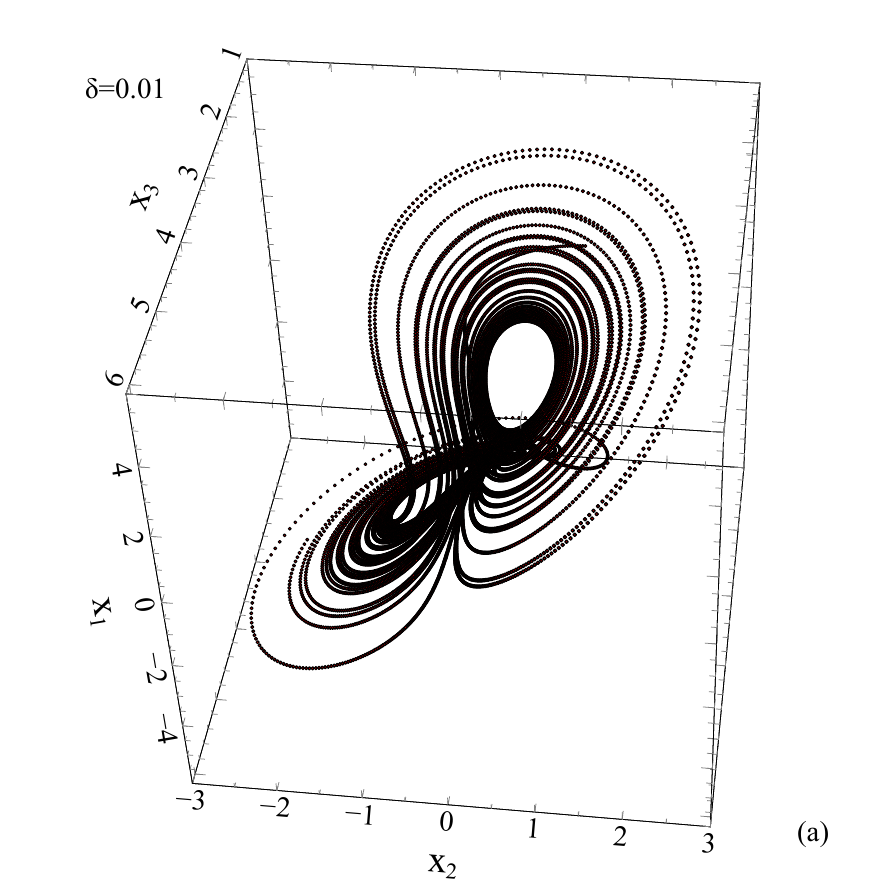}
	\includegraphics[height=6cm,width=7cm]{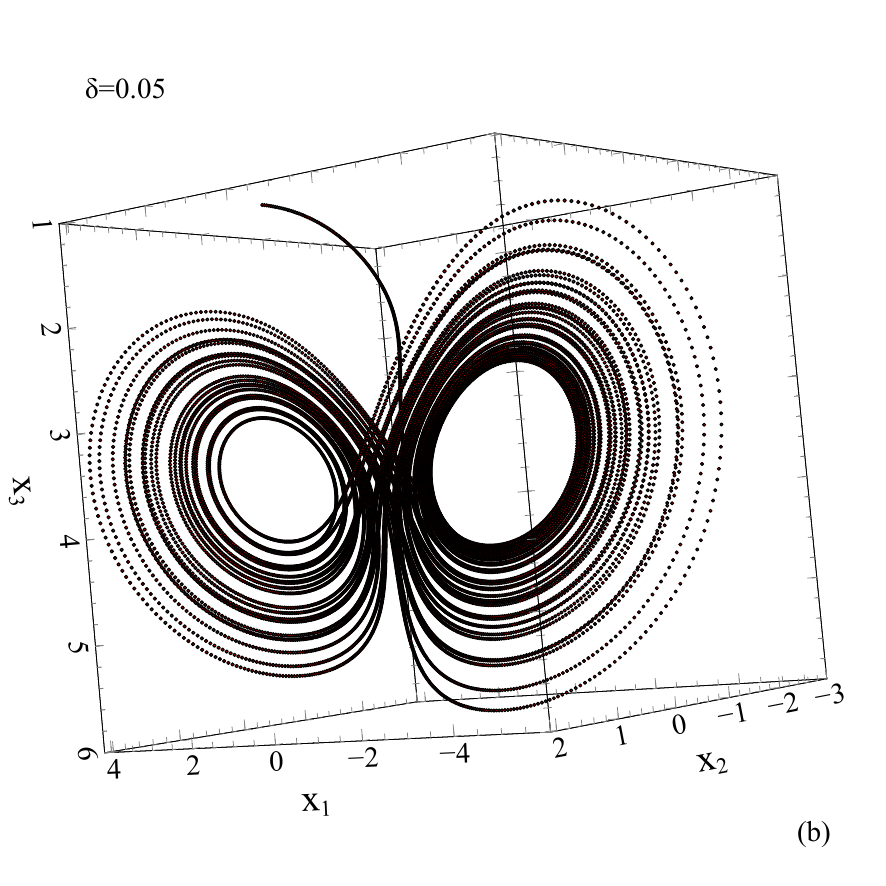}
	\includegraphics[height=6cm,width=7cm]{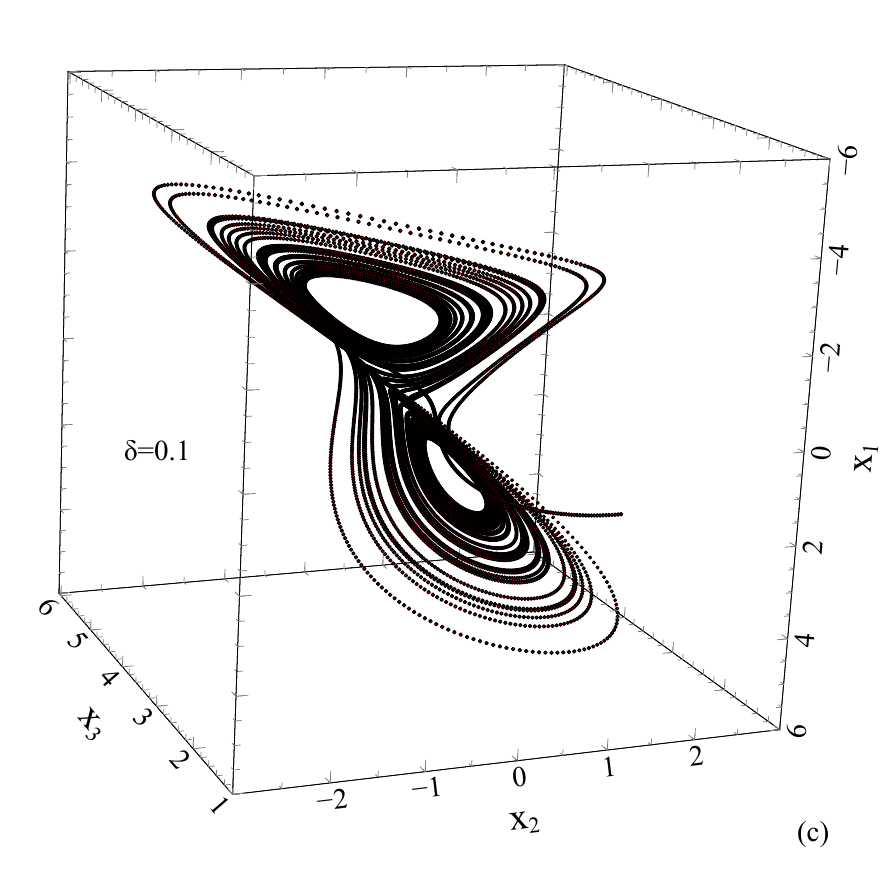}
	\includegraphics[height=6cm,width=7cm]{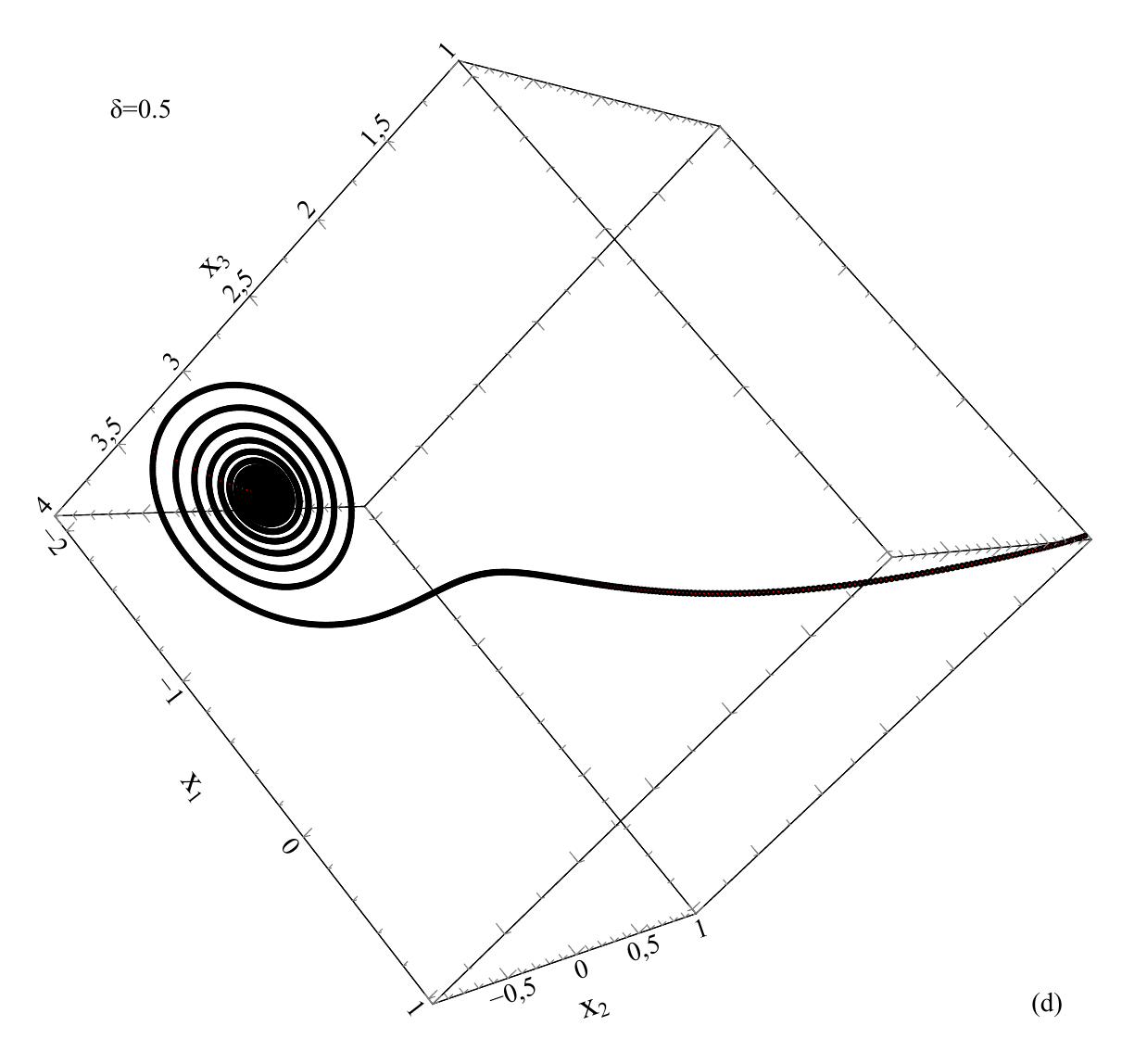}
	\caption{Attractors in $x_{1}-x_{2}-x_{3}$ phase-space for the different control parameters $\delta$.}
	\label{a-figs}
\end{figure*}

The integral terms in the expression Eq.(\ref{Aydiner2}) represent the energy storage capacity in systems. Integral expressions do not contain non-linear terms. Therefore, contributions from these terms can be considered as constant contributions in the numerical solution. For simplicity, we define these integrals as $\delta_{1,2}$, where $\delta_{1} = \int_{0}^{\tau} x_{1}(t) dt$ and $\delta_{2} = \int_{0}^{\tau} x_{2}(t) dt$ we can simplify the solution. We can think of $\delta_{1,2}$ as control parameters that measure the energy storage capacity. In this case, rather than getting lost in the details of the numerical solution, it would be appropriate to include the behaviour of the system depending on the parameter $\delta_{1,2}$ for a fixed value $q$. Because, as can be seen in Fig.\ref{Lyap-1}, the system behaves chaotic for all $q$ values. However, the chaotic dynamics in the system can be controlled depending on the energy storage capacity.

In order to maintain simplicity in the numerical solution, we will take $\delta=\delta_{1}=\delta_{2}$ and solve Eq.(\ref{Aydiner2}) numerically for different values $\delta$. Using the Runga-Kutta method and the linearized algorithm $\delta$ values are given in Fig.\ref{a-figs} for the same $q$ values.

\begin{figure*} [ht]
	\centering
	\includegraphics[height=6cm,width=7cm]{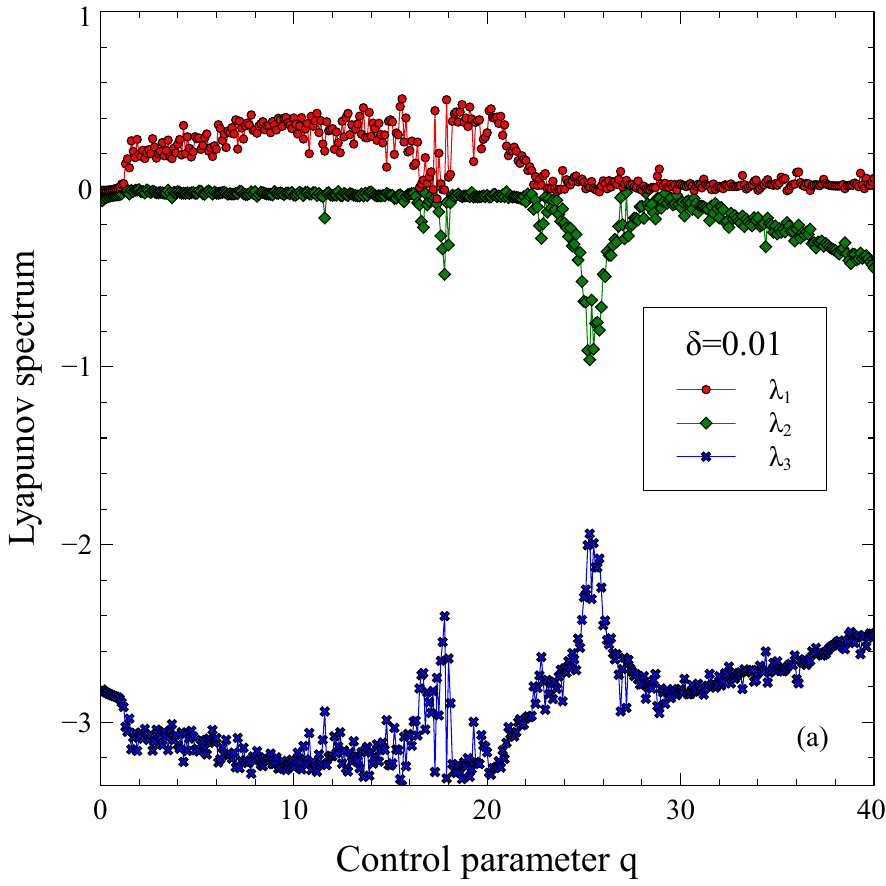}
	\includegraphics[height=6cm,width=7cm]{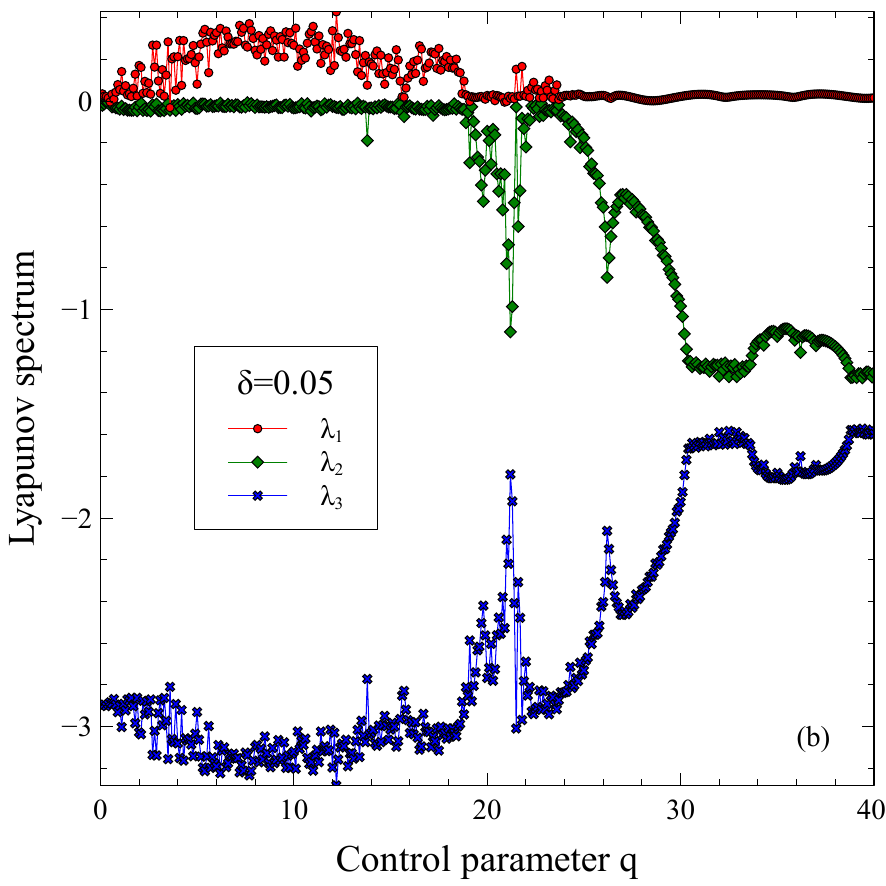}
	\includegraphics[height=6cm,width=7cm]{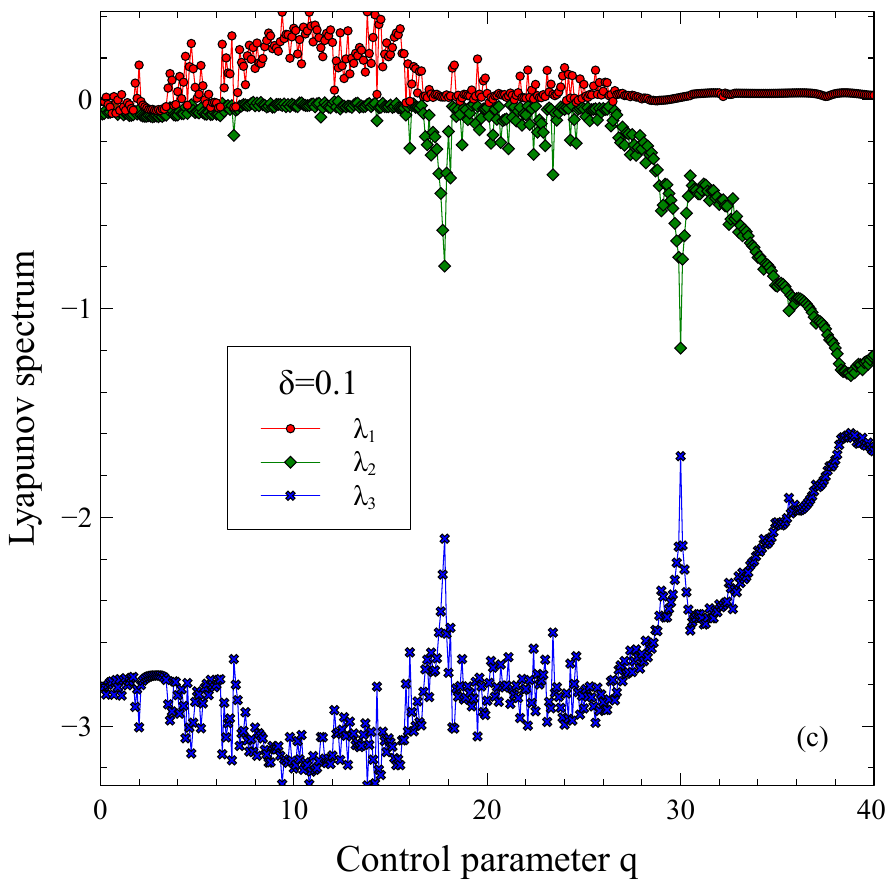}
	\includegraphics[height=6cm,width=7cm]{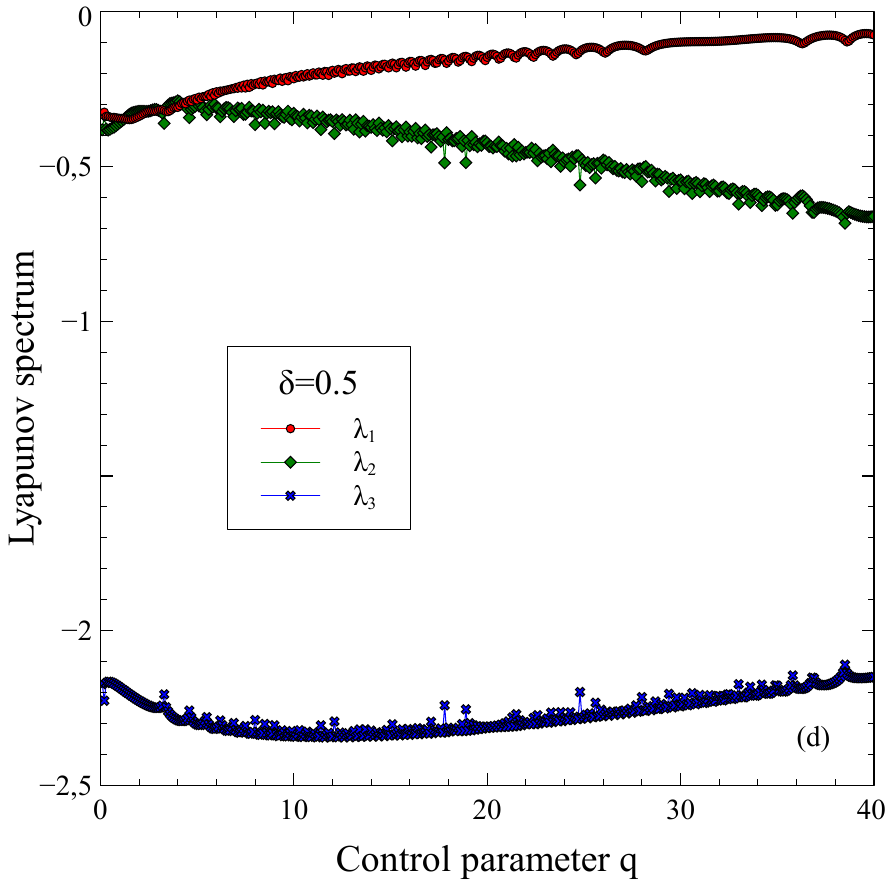}
	\caption{Three Lyapunov exponents for various $\delta$ parameter.}
	\label{Lyap-2}
\end{figure*}

Figs.\ref{a-figs} (a)-(d) for four different delta values $\delta =0.01$, $\delta =0.05$, $\delta =0.1$, and $\delta =0.5$ are plotted. As can be seen in Fig.\ref{a-figs} (a)-(c), the existence of chaotic attractors in the phase space is observed for small values of $\delta$. But for $\delta =0.5$ the chaotic attractors disappear. We did not look for a critical value for $\delta$ here. However, we only wanted to show that for large values of the $\delta$ parameter, chaotic dynamics will disappear, and the system will go to a periodic or aperiodic solution. Here, $\delta$ acts as a control parameter.

We notice that the phase-space solution in Figs.\ref{a-figs} (a)-(d) are plotted for only same arbitrary $q$ value.
However, we see that the character of the phase-space solutions in Figs.\ref{a-figs} (a)-(d) do not change for the different $q$ values. As seen form Figs.\ref{a-figs}(a)-(c) that the chaotic attractor is located around two fixed point like Lorenz attractors in the $x_{1}-x_{2}-x_{3}$ phase space.  Similarly, one can see that from Figs.\ref{a-figs} (a)-(c) the trajectories within the attractors do not intersect.  On the other hand, it should also be noted that when the integral contribution in Eq.(\ref{Aydiner2}) grows, the system ceases to be chaotic. Because the integral represents the work done by the volume change on the system. This means an energy storage capacity. As seen in Figs.\ref{a-figs} (d), when this capacity grows, the interaction dynamics of the systems cease to be chaotic.

Similarly, to obtain further information about dynamics we calculated the Lyapunov exponents for the same values $\delta$ and gave them in Fig.\ref{Lyap-2}. As can be seen in Figs.\ref{Lyap-2} (a) and (c), there is at least one positive Lyapunov exponent in the spectrum for $\delta =0.01$, $\delta =0.05$ and $\delta =0.1$ For $\delta =0.5$, in Fig.\ref{Lyap-2} (a), all exponents do not take negative values. As will be remembered, to say that the dynamics of the system are chaotic, there must be at least one positive exponent in the Lyapunov spectrum.

We would like to point out another important difference. For example, for conditions $\delta Q \ne 0$, $\delta N=0$ and $dV=0$, only one chaotic attractor appears in the phase space, while for case $\delta Q \ne 0$, $\delta N\ne 0 $ it is seen that it is a chaotic attractor in the condition and $dV \ne0$. The reason for this can be expressed as the system of equations given in Eq.(\ref{Aydiner2}) has a more complex structure.

Another important point is that $\delta=\delta_{1}=\delta_{2}$ was chosen. Instead, if different values of $\delta_{1}$ and $\delta_{2}$ had been studied, two control parameters would have been included in the system. However, it is obvious that for small values $\delta_{1}$ and $\delta_{2}$, the system given by Eq.(\ref{Aydiner2}) will produce chaotic dynamics. Another point is that the expressions $\delta_{1}$ and $\delta_{2}$ could be written as a function of time. But it is clear that, in any case, the quantitative analysis will not change.

\section{Discussion} \label{disc}

In our previous work, we considered two thermodynamic systems interacting under the condition \cite{Aydiner_2024} $\delta Q \ne 0$, $\delta N \ne 0$, $dV=0$. We defined a new interaction scheme. We proved the validity of this interaction scheme by starting from the law of conservation of energy. From these results, we suggested that the energy interaction of dark matter and dark energy in particular should be chaotic. 

In this current study, we have shown that the interaction dynamics of thermodynamic systems interacting under conditions $\delta Q \ne 0$, $\delta N=0$, $dV=0$ and $\delta Q \ne 0$, $\delta N\ne 0$, $dV \ne0$ are chaotic. These new findings show that the proposed new law may be universal for all interacting thermodynamic systems. However, although it is not based on a experimental and theoretical proof, it is possible to assume that the proposed law may be also valid for particle-particle interactions. We will return to this topic.

As can be seen from Figs.\ref{phase-1}, \ref{phase-2} and \ref{phase-3}, in the case of $\delta N=0$, a single chaotic attractor is formed. The formation of a single attractor does not show that the dynamics are not chaotic. When we look at the figures, we can see that the orbits are progressing around a fixed point without intersecting each other. When we look at Fig.\ref{Lyap-1} giving the Lyapunov exponents, it shows that $\lambda_1$ (red curve) and $\lambda_2$ (green curve) take positive values, which shows that the Eq.(\ref{Aydiner1}) has chaotic dynamics. However, as can be seen from Figs.\ref{a-figs} (a)-(c), in the case of Eq.(\ref{Aydiner2}), namely $\delta N \neq 0$, two chaotic attractors are formed, similar to Lorenz's butterfly curve.

Now, we want to get back to the noteworthy problems: \textit{Why do we expect the dynamics of interacting systems to be chaotic? What could be the physical and philosophical equivalent of this question?} These two questions are important. The answers to these questions are available in the proposed law and in the given definition. This law and definition clearly imply that: \textit{"Chaos minimizes the joint action".} To understand this, let us recall the trajectories in the Lorenz butterfly attractors. The Lorenz equation \cite{Lorenz_1963} is formed by writing three differential equations containing three first-order time derivatives in a coupled form. If these three equations were not coupled, that is, if they were in separable form, each equation would have separate solutions and separate trajectories. However, the coupled form of the equation produces a single common trajectory that minimizes all three variables simultaneously, which is the shortest possible path for the chosen parameters. It is not one of the shortest paths, but the only shortest path. This shows us that chaos produces the shortest path. This is why it leads to a correct conceptual description of this process. Once again, this definition implies the existence of a new law of physics, which is introduced here.

On the other hand, let us point out that there is no definition of chaos in the literature \cite{Gleick_1988,Weisstein}. Chaos is generally called a process and is well known for its three basic features. These are; i) it arises from non-linear dynamics, ii) it shows sensitive dependence on the initial condition, and iii) its evolution over time is unpredictable \cite{May1976simple,Holmgren1994,Elaydi1999,Devaney1986,Strogatz1994,Hilborn1994,Abarbanel1993,Wiggins1990,Tabor1989,Rasband1990,Ruelle1971,Sprott2003chaos,Martelli_1998,Batterman_1993,Banks_1992,Ott_1993,Lorenz_1996,Hao_1984,Smith_1998}. Even so, one can find chaos definitions in the literature.
For instance; the term chaos has first been used in 1975 by Li and York in their paper \cite{Yorke1975}: "Period three implies chaos". Wiggins \cite{Wiggins1990} define chaos as: "A dynamical system displaying sensitive dependence on initial conditions on a closed invariant set (which consists of more than one orbit) will be called chaotic". Tabor \cite{Tabor1989}: "By a chaotic solution to a deterministic equation we mean a solution whose outcome is very sensitive to initial conditions (i.e., small changes in initial conditions lead to great differences in outcome) and whose evolution through phase space appears to be quite random." Rasband \cite{Rasband1990}: "The very use of the word 'chaos' implies some observation of a system, perhaps through measurement, and that these observations or measurements vary unpredictably. We often say observations are chaotic when there is no discernible regularity or order". Prigogine defined chaos as 'order emerging from disorder' in his book titled "Order Out of Chaos" \cite{Prigogine_1984}. It is possible to multiply such definitions.
In contrast to these definitions, our 
chaos  definition is based on the minimization of action and is related to the \textit{Least Action Principle} (LAP) \cite{Maupertuis_1744}. We will return to this issue below.

As is known, LAP \cite{Euler_1767tb,Newton_1687tb,Lagrange_1772tb}
defines the shortest possible path of an object in a force field. LAP is the most fundamental principle of physics. All fundamental laws of physics can be expressed in terms of a LAP. Another important issue is that LAP preserves causality. However, there are physical processes where the validity of LAP is not clearly clear. For example; LAP is necessary but might not sufficient for the stochastic processes, non-equilibrium and chaotic systems. So, it can be seen in the literature that new action principles have been recently proposed for stochastic, non-equilibrium, and chaotic systems \cite{Wang_2005,Wang_2014,Tongling_2013,Magnenet_2020,Bufalo_2021}.

Here, like LAP, chaos also defines the shortest path, that is, the minimum action. This is the physical and philosophical meaning of the law and the definition that we propose. If we make what we say more formal, we can introduce a new idea: \textit{"Non-linear nature minimizes own action with chaos"}. This inference tells us of the existence of a new principle. Its physical and philosophical equivalent is \textit{"Chaotic Action Principle"} (CAP).

CAP reveals the physical and philosophical basis of the behaviour of chaotic systems. Thanks to CAP, we can get a better idea of how nature minimizes its action. On the other hand, CAP can help us understand the relationship between chaotic dynamics and geometric representations of fractal structures that we encounter in nature. Moreover, CAP may explain the conservation of causality in chaotic organized nature. This is the meaning and implication of CAP. In a non-linear nature, causality may be preserved by CAP.
This new principle does not reject LAP. In this sense, CAP is a generalization of LAP. 

Returning to the chaotic nature of particle interactions, we have several motivations here. Our first and most important motivation is the chaotic nature of the three-body problem \cite{Poincare_1890tb}. Our second motivation is the macroscopic nature of composite particles. Our third motivation is the fractal structure of nature. Another motivation is that the non-linear nature of the dynamics of the interactions of elementary particles defined in the standard model is not yet fully understood. Our final motivation is that the dynamics of all micro systems should also obey the CAP.

If we remember that the three-body problem has a very complex dynamics \cite{Poincare_1890tb}. There are no closed-form solutions to the three-body problem, and the dynamics is clearly chaotic. Although there are very special analytical solutions \cite{Euler_1767tb,Lagrange_1772tb,Jacobi1843,Delaunay1860,Hill1886,Hill1878} and numerous numerical solutions \cite{LI_2019tb,Hristov_2024tb} to this problem, the dynamics is chaotic. This problem is the most important physical phenomenon that shows how nature works and supports our theses here. As for the second motivation, in nature, particles that are composite and sufficiently small are actually thermodynamic systems. However, these systems have geometric forms that cannot be idealized in an imaginary way. Therefore, such binary particle systems naturally turn into a three-body problem due to their geometric and centre-of-mass asymmetries. As for the third motivation, nature and the universe have fractal geometry from micro- to macro-scale. This fractal formation can be interpreted as a manifestation of chaotic dynamics. Now let us turn to the last motivation, the interactions of elementary particles.

The law we propose can be expected to be valid at the level of elementary particle interactions, namely fermions and bosons. There are two reasons for thinking this. First, the non-linearity in nature is expected to be valid not only for macroscopic particles and systems but also in the microscopic world, namely at the level of elementary particle interactions. Second, the energy-momentum exchanges of elementary particles can be modelled as in thermodynamic systems. On the other hand, studies in the literature that imply that the interactions of elementary particles can be chaotic \cite{Beck_2002,Beck_2008,Schafer_2011,Groote_2014}.

Now, we will discuss on relation between CAP and complexity over the new definition of chaos. As is known, there is no conceptual definition in the literature for the question "What is a complex system?" like the question "What is chaos?" complex systems have also been tried to be defined in the literature through their features \cite{Ladyman_2013,Estrada_2024}. However, there is no conceptual definition that everyone agrees on. However, we believe that it is possible to give a conceptual definition of "complexity" with the definition of chaos and self-organization that we have given here and in the previous study \cite{Aydiner_2024} and with the help of new conceptual tools such as CAP. Although there is no formal proof yet, we will propose a new definition of complex system as: \textit{"A complex system is a self-organized system formed by the subgroups of self-organized systems based on the CAP".} Now, we will not go into the details of this definition here. One might find this definition useful. For example; when we go through this new conceptual definition of a complex system, it will be seen that it is easier to distinguish between self-organized systems and complex systems. For example, while a piece of rock is in the category of self-organized systems, it will be more understandable to evaluate a human brain or Earth's global climate as being in the category of complex systems. Finally, these discussions may provide clues that \textit{"living systems such as humans, animals and plants can be called complex systems that can replicate themselves"}.
We would like to point out that we will discuss these new definitions and the literature on this subject in detail in the future.

\section{Conclusion} \label{conc}

In our previous work \cite{Aydiner_2024}, we showed that the interaction of two thermodynamic systems under the condition $\delta Q \ne 0$, $\delta N \ne 0$, $dV=0$ is chaotic. In the present study, we considered thermodynamic systems that interact with each other under conditions $\delta Q \ne 0$, $\delta N=0$, $dV=0$ and $\delta Q \ne 0$, $\delta N\ne 0$, $dV \ne0$. Using the interaction schema presented in Ref.\cite{Aydiner_2024}, we obtained Eq.(\ref{Aydiner1}) for the conditions $\delta Q \ne 0$, $\delta N=0$, $dV=0$ and Eq.(\ref{Aydiner2}) for the conditions $\delta Q \ne 0$, $\delta N\ne 0$, $dV \ne0$. We obtained numerical solutions by solving these equations with FORTRAN 90.

It is understood from the Lyapunov spectrum given in Fig.\ref{Lyap-1} that the interaction dynamics of two closed thermodynamic systems that interact only through heat transfer is chaotic. Similarly, by including heat exchange and work done on the system in the equation, it has been shown that the interactions of the open thermodynamic system given in Eq.(\ref{Aydiner1}) will be chaotic. The results obtained here are in agreement with the results obtained in Ref.\cite{Aydiner_2024}. In other words, the results obtained confirm the universality of the new physics law we proposed.

Additionally, we should point out that Case II gives slightly newer, more interesting results. However, the integral terms in the equation system Eq.(\ref{Aydiner2}) work as control parameters and can be used to remove the dynamics of the system from chaos. This parameter will be possible to determine the chaotic threshold of the system. Large values of this parameter may suppress other interactions in the interacting system. Such interactive systems may require further investigation.

This study is complementary to the work presented in Ref.\cite{Aydiner_2024}. The results obtained in the previous study and in this study suggest that the dynamics of interacting thermodynamic systems will be chaotic. In other words, the results obtained support the completeness and universality of the proposed physical law. Finally, we point out that, leaving the physics aside, Eq.(\ref{Aydiner1}) and Eq.(\ref{Aydiner2}) are mathematically interesting in themselves. The fact that equations in this form produce chaos is a meaningful and important result in itself.

Particularly, in Section \ref{disc}, we focused on the theoretical and philosophical background of the new physics law and chaos definition. We discussed the connection between chaos and the principle of minimum action. It is first time, we introduced a new action principle: CAP. Then, we argued that the principle of chaotic action should be valid for all of nature and emphasized that the proposed law of physics should be valid for the interaction of elementary particles. We evaluated that the chaotic action principle we proposed could provide new ideas for understanding and defining nature and especially complex systems, and we gave the definition of complex systems. 

In summary, this study is an effort to understand the dynamics of nature and create a new perspective. The models, definitions and concepts we present in this study are open to discussion. In future studies, we will try to develop the topics and concepts we have discussed here and our suggestions.

\section*{Acknowledgment}
I am deeply grateful to the esteemed referee who reviewed my article. This work has greatly benefited from the referee’s insightful contributions and pioneering wisdom, and it has become what it is thanks to their thoughtful guidance.




\bibliographystyle{elsarticle-num}
\bibliography{Ankara}







\end{document}